\documentclass[aps,prc,twocolumn,floatfix,showpacs,preprintnumbers,amsmath,amssymb,nofootinbib,groupedaddress]{revtex4-1}
\usepackage{color}
\usepackage{graphicx}
\usepackage{dcolumn}    % Align table columns on decimal point
\usepackage{multirow}
\usepackage{bm}         % bold math
\usepackage{sidecap}
\usepackage{hyperref}   % use for hypertext links, including those to external documents and URLs
\usepackage{xcolor}
\usepackage{amsmath}
\usepackage{color,colortbl}
\usepackage{booktabs}  
\usepackage{hyperref} 
\definecolor{dark-red}{rgb}{0.,0.,0}
\definecolor{dark-blue}{rgb}{0.,0.,1}
\definecolor{medium-blue}{rgb}{0,0,1}
\definecolor{gray}{rgb}{0.85,0.85,0.85}

\hypersetup{
    colorlinks, linkcolor={dark-red},
    citecolor={dark-blue}, urlcolor={medium-blue}
}

\begin{document}
% >>>>>>>>>>>>>>>>>>>>>>>>>>>>>>>>>>>>>>>>>>>>>>>>>>>>>>>>>>>>>>>>>>>>

\title{Low-energy quadrupole states in neutron-rich tin nuclei}

\author{E. Y{\"{u}}ksel\textsuperscript{1,2}}
\email{eyuksel@yildiz.edu.tr}
\author{G. Col\`o\textsuperscript{3,4}}
\email{gianluca.colo@mi.infn.it}
\author{E. Khan\textsuperscript{5}}
\email{ khan@ipno.in2p3.fr}
\author{Y.F. Niu\textsuperscript{6}}
\email{nyfster@gmail.com}

\affiliation{\textsuperscript{1}  Department of Physics, Faculty of Science, University of Zagreb, Bijeni\v{c}ka c. 32, 10000 Zagreb, Croatia\\
\textsuperscript{2} Department of Physics, Yildiz Technical University, 34220
Esenler, Istanbul, Turkey\\
             \textsuperscript{3} Dipartimento di Fisica, Universit\`a degli 
             Studi di Milano, via Celoria 16, I-20133 Milano, Italy\\	
							\textsuperscript{4} INFN, Sezione di Milano, Via Celoria 16, 20133 Milano, Italy \\				
               \textsuperscript{5} Institut de Physique Nucl\'eaire, Universit\'e Paris-Sud, IN2P3-CNRS, \\
							Universite Paris-Saclay, F-91406 Orsay Cedex, France \\
               \textsuperscript{6} ELI-NP, Horia Hulubei National Institute for Physics and Nuclear Engineering, 30 Reactorului Street, RO-077125, Bucharest-Magurele, Romania}

\date{\today} 

% >>>>>>>>>>>>>>>>>>>>>>>>>>>>>>>>>>>>>>>>>>>>>>>>>>>>>>>>>>>>>>>>>>>>
% ABSTRACT, PACS.
%
\begin{abstract}
We present a study on the isoscalar quadrupole strength in tin nuclei, focusing mainly on the low-energy region. The calculations are performed using the Skyrme type energy density functionals within the fully self-consistent quasiparticle random phase approximation, allowing for a good description of the experimental data for the first 2$^+$ state and the isoscalar giant quadrupole resonance.
It is found that the first $2^+$ state and the low-energy quadrupole states between 3 and 6 MeV display an opposite behavior with increasing neutron number.
While the strength of the first $2^+$ state decreases, some excited states start to accumulate between 3 and 6 MeV, and increase their strength with increasing neutron number. 
This low-energy region between 3 and 6 MeV is quite sensitive to the changes in the shell structure with increasing neutron number. 
In particular, between $^{116}$Sn and $^{132}$Sn, the filling of the neutron
orbitals with large values of $j$, has an important impact on the low-energy region. Our analysis
shows that the low-energy states have a non-collective character, except the first 2$^+$ state. In addition, the states in
the low-energy region above 5 MeV display an interesting pattern:
with the increase of the neutron number, their strength increases 
and their nature changes, namely they switch from proton excitations 
to neutron-dominated one.
We conclude that the low-energy quadrupole states between 3 and 6 MeV 
can provide information about the shell evolution in open-shell nuclei.
\end{abstract}

%  21.10.Gv Nucleon distributions and halo features
%  21.10.Pc Single-particle levels and strength functions
%  21.10.Re Collective levels
%  21.60.Ev Collective models
%  21.60.Jz Nuclear Density Functional Theory and extensions (includes Hartree-Fock and random-phase approximations)
%  21.65.Ef Symmetry energy
%  24.30.Cz Giant resonances 
%  24.30.Gd Other resonances 
%  25.20.-x Photonuclear reactions
%  25.60.-t Reactions induced by unstable nuclei
%  25.40.Kv Nucleon-ind,vre reactions - Charge-exchange reactions 
%  25.55.Kr 3H-, 3He-, and 4He-induced reactions - Charge-exchange reactions 
%  23.40.Hc Beta decay; double beta decay; electron and muon capture - Relation with nuclear matrix elements and nuclear structure 

%\keywords{}

\maketitle
\section{Introduction}
Giant resonances (GR) are one of the most important tools in order to probe 
the properties of nuclei, nuclear matter, and the nuclear Equation of 
State. They already provided a wealth of structure information
\cite{paar00,hara01}. In recent decades, the development of new experimental facilities, 
allowed to reach part of the regions of the nuclear chart that extends out of the stability valley.
Therefore, theoretical investigations have also moved along the
same line in order to understand the structure and properties of exotic nuclei 
with unusual neutron-to-proton ratios. One of the most 
interesting property of exotic nuclei is the formation of the 
new low-energy dipole states which increase their strength with the neutron number along 
an isotopic chain.

These low-energy states are generally found 
below the giant dipole resonance region, that is, between 8 and 12 MeV 
\cite{paar00,paar09,vre12,yuk12,roc12,lan09,sav13,bra16,wie09,krum15}. It has been shown that 
the low-energy dipole states play a relevant role 
in astrophysical processes \cite{gori02,lit09}, and 
it has been suggested that they can also provide information 
about the structure of exotic nuclei \cite{pie06,klim07}. 
While some authors have employed the term ''Pigmy Dipole Resonance'' 
(PDR) in this context, the appearance of a resonance peak, if any, 
is certainly nucleus-dependent and its interpretation is probably model-dependent.

Apart from the low-energy dipole states, the formation of the low-energy quadrupole states was also addressed in several works. Among such first theoretical works are the calculations 
performed using the well known Random Phase Approximation (RPA) for the $^{28}$O nucleus, which is predicted as bound, although unbound experimentally \cite{tar97}. In this work, the low-energy excitations were referred 
as giant neutron modes \cite{yok95}. 
The quadrupole response in $^{28}$O nucleus has also been
studied using the continuum RPA \cite{ham97}. In this case, the 
emergence of a low-energy excitation is clearly a threshold effect and 
not a collective one.
Then, the distribution of the low-energy quadrupole strength was investigated for Nickel isotopes, 
using both the shell model and QRPA with separable interactions \cite{lan03}.
The authors predicted some fragmented low-energy quadrupole strength 
above 4 MeV with increasing neutron number. Using RPA, an increase in the low-energy quadrupole strength was also obtained in the selected tin nuclei with increasing neutron number \cite{lan09}.
Recently, the formation of low-energy quadrupole states was also predicted on Sn nuclei using the quasiparticle phonon model (QPM) on top of Hartree-Fock-Bogoliubov (HFB) calculations \cite{tso11}. In this work, a group of low-energy quadrupole states 
was found between 2 and 4 MeV, when increasing the neutron number, 
and referred as pygmy quadrupole resonance due to their different 
character than the first $2^+$ state and 
the isoscalar giant quadrupole resonance (ISGQR).

The first experimental signatures for the so-called pygmy quadrupole resonance states 
were obtained in the ($^{17}$O,$^{17}$O'$\gamma$) reaction 
at 340 MeV \cite{pel15} in the case of the $^{124}$Sn nucleus, 
where a group of $2^+$ states were measured between 2 and 5 MeV and interpreted as a quadrupole type oscillation of the neutron skin. 
Recently, the low-energy quadrupole states were also obtained using ($\alpha,\alpha'\gamma$) and ($\gamma,\gamma'$) 
experiments in $^{124}$Sn \cite{spi16}, complementing the experimental results given in Ref. \cite{pel15}. In a recent review about the pygmy dipole states, possible formation of the pygmy quadrupole states in neutron-rich nuclei was also pointed out \cite{bra16}.

Considering the recent theoretical and experimental results, 
it would be interesting to investigate the sensitivity of the low-energy quadrupole states to the neutron excess. 
Although considerable 
amount of work has been devoted to studying either the giant 
quadrupole resonance or the first $2^+$ states of nuclei 
\cite{ter02,fle04,ter06,an06,sev08,scam13}, there is not, to (the best of) our knowledge, a detailed study addressing
if and why some different kind of low-energy quadrupole strength
may develop as a function of the neutron number. 
Therefore, in this work, we aim to explore the evolution of the low-energy quadrupole states by increasing the neutron number, using 
the quasiparticle random phase approximation (QRPA) with Skyrme 
type energy density functionals. The calculations are fully self-consistent: time-odd terms, spin-orbit and Coulomb are included. QRPA is a well known tool to study 
the nuclear response, so that it can
help to clarify the detailed structure of the low-energy states below the 
giant resonance region. Although QRPA does not include more complex configurations 
than two quasi-particles \cite{niu16}, predictions about the evolution and nature of the states in the low-energy 
region with increasing neutron number can be nevertheless explored. As mentioned above, the low-energy region between 2 and 5 MeV is named as pygmy quadrupole resonance region in Refs. [20–22]. However, nature of these low-energy states is
not clear yet as we will discuss below. In the present work, this region between the lowest $2^+$ state and ISGQR region is hereafter named as the low-energy region, rather than the pygmy quadrupole resonance region.

The paper is organized as follows. In Section \ref{theo}, we briefly summarize the theoretical framework and the QRPA. In Section \ref{res},
some ground state properties are discussed briefly, and the isoscalar quadrupole response results are given for the Sn isotopic chain. Especially, we focus on the low-energy quadrupole states in order to clarify their nature and behavior with increasing neutron number. The configurations of the excited states and transition densities are also analyzed. Finally, the summary and conclusions are given in Sec. \ref{finito}. 

\section{Theoretical Framework} \label{theo}
In the present work, the ground state properties of nuclei are calculated within the Hartree-Fock BCS framework, using Skyrme type SkM* \cite{bar82}, SGII \cite{SGII} and SLy5 \cite{SLY5} interactions. In the calculations, a zero-range density-dependent pairing interaction is used \cite{ber91}
\begin{equation}
V_{pair}(\textbf{r}_{1},\textbf{r}_{2})=V_{0}\left[1-\eta\left(\frac{\rho(\textbf{r})}{\rho_{0}}\right)\right]\delta(\textbf{r}_{1}-\textbf{r}_{2}),
\end{equation}
where $\rho_{0}=$0.16 fm$^{-3}$ is the nuclear saturation density and 
$\rho(\textbf{r})$ is the particle density. The parameter $\eta$ can be set between zero and one, i.e. the volume and surface type pairing interactions, respectively. Our procedure to constrain the parameters of the density-dependent pairing interaction is to consider the experimental values of both the neutron pairing gap according to the three-point formula \cite{sat98} and the first $2^+$ state energies and transition probabilities in Sn isotopes. We find that the SkM* interaction with volume type pairing, which is a specific case of the density-dependent pairing interaction, reproduces the best the experimental data for the first $2^+$ state energies and transition probabilities in the Sn isotopes as it will be discussed in Sections \ref{1} and \ref{a1}. Therefore, in the present work, we shall continue our analysis with volume type pairing interaction. In the calculations, the pairing strength ($V_{0}$) is taken as 210.0, 230.0 and 250.0 MeV.fm$^{3}$ for SkM*, SGII and SLy5 interactions, respectively.

In this work, the QRPA calculations are performed on top of the Hartree-Fock BCS approach. Since the QRPA method is a well known tool for the calculation of the multipole response of nuclei (see Refs. \cite{ring80,suho07,li08,bai14,paar03,niu17}), we only mention the main equations in this section.
The QRPA matrix is given by
\begin{equation}
\left( { \begin{array}{cc}
 A & B  \\
 -B^{*} & -A^{*}  \\
 \end{array} } \right)
 \left( {\begin{array}{cc}
 X^{\nu}   \\
 Y^{\nu}  \\
 \end{array} } \right)
 =E_{\nu}
  \left( {\begin{array}{cc}
 X^{\nu}   \\
 Y^{\nu}  \\
\end{array} } \right).\end{equation}
The A and B matrices are defined by
\begin{gather}
\begin{aligned}
\label{eq:a}
A_{abcd}&=(E_{a}+E_{b})+(u_{a}u_{b}u_{c}u_{d}+v_{a}v_{b}v_{c}v_{d})V^{\text{pp}}_{abcd} \\
&+N_{ab}(J)N_{cd}(J)\big[(u_{a}v_{b}u_{c}v_{d}+v_{a}u_{b}v_{c}u_{d})V^{\text{ph}}_{a\bar{d}\bar{b}c} \\
&-(-1)^{j_{c}+j_{d}+J}(u_{a}v_{b}v_{c}u_{d}+v_{a}u_{b}u_{c}v_{d})V^{\text{ph}}_{a\bar{c}\bar{b}d}\big],
\end{aligned} \\
\begin{aligned}
\label{eq:b}
B_{abcd}&=-(u_{a}u_{b}v_{c}v_{d}+v_{a}v_{b}u_{c}u_{d})V^{\text{pp}}_{ab\bar{c}\bar{d}} \\
&+N_{ab}(J)N_{cd}(J)\big[(u_{a}v_{b}v_{c}u_{d}+v_{a}u_{b}u_{c}v_{d})V^{\text{ph}}_{ad\bar{b}\bar{c}} \\
&-(-1)^{j_{c}+j_{d}+J}(u_{a}v_{b}u_{c}v_{d}+v_{a}u_{b}v_{c}u_{d})V^{\text{ph}}_{ac\bar{b}\bar{d}}\big],\\
\end{aligned}
\end{gather}
where $E_{a(b)}$ is the quasiparticle (q.p.) energy of the states, $u$ and $v$ are the BCS occupation factors, and $N_{ab(cd)}$ is the normalization constant. In addition, $V^{\text{pp}}_{abcd}$ and $V^{\text{ph}}_{adbc}$ represent the particle-particle (pp) and particle-hole (ph) residual interactions, respectively.
For a given excited state $ E_{\nu}$, the contribution of the proton and neutron quasiparticle configurations
is determined by the QRPA amplitudes,
\begin{equation}
A_{ab}=|X_{ab}^{\nu}|^{2}-|Y_{ab}^{\nu}|^{2},
\label{Aab}
\end{equation} 
and the normalization condition can be written as
\begin{equation}
\sum_{a\geq b}A_{ab}=1.
\label{aaa}
\end{equation}
The reduced transition probability for any operator $\hat{F}_{J}$ is also given by
\begin{equation}
\begin{split}
B(EJ,\widetilde0\rightarrow \nu)=&\biggl|\sum_{c\geq d}b_{cd}(EJ)\biggr|^{2}\\
=&\biggl|\sum_{c\geq d}(X_{cd}^{\nu}
+Y_{cd}^{\nu})(v_{c}u_{d}+u_{c}v_{d})\langle c ||\hat{F}_{J}||d\rangle\biggr|^{2},
\label{bel}
\end{split}
\end{equation}
where $|\nu\rangle$  is the excited state and $|\widetilde0\rangle$ is the correlated QRPA ground state. The energy weighted moments $m_{1}$ and $m_{0}$ are defined using
\begin{equation}
m_{k}=\sum_{\nu}B(EJ,\widetilde0\rightarrow \nu)E_{\nu}^{k}.
\end{equation}   
In this work, the continuum is discretized inside a spherical
box of 20 fm with 0.1 fm mesh size. We use a large
quasiparticle energy cut-off ($E_{cut}=100$ MeV) in the QRPA calculations in order to satisfy the Energy
Weighted Sum Rule (EWSR). The discrete QRPA spectra are averaged with a
Lorentzian having width $\Gamma=1$ MeV. The spherical symmetry is assumed in the calculations, which is appropriate to study the low-energy quadrupole states in Sn isotopic chain.

\section{results} \label{res}
In this section, the isoscalar quadrupole response in superfluid tin nuclei is investigated within the Skyrme energy density functional framework, using the QRPA. A special emphasis is given to the low-energy region of the quadrupole strength in order to investigate and clarify the possible enhancement of this strength with increasing neutron number.
\subsection{Evolution of pairing correlations and of the single
(quasi)-particle states}\label{1}
Before discussing the evolution of the low-energy quadrupole strength in tin nuclei, it is also useful to investigate the 
changes in the pairing correlations and single(quasi)-particle states with increasing neutron number. Therefore, in this part, we first give a brief overview of these changes relevant for discussions of Sec. \ref{a1}.
\begin{figure}[!ht]
 \begin{center}
\includegraphics[width=0.95\linewidth,clip=true]{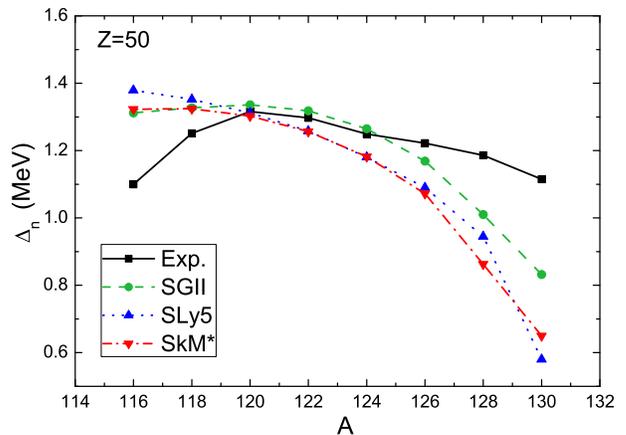}
 \end{center}
 \caption{Calculated mean value of the neutron pairing gap using Hartree-Fock BCS approach and Skyrme type SGII, SLy5 and SkM* interactions compared with the experimental values.} 
  \label{mean}
\end{figure}

It is well known that the first $2^+$ state is rather sensitive to pairing 
\cite{scam13}, while the impact of pairing on the ISGQR is weak. In order to reach a more general understanding, the mean value of the neutron pairing gap $\Delta_{n}$ is calculated using Hartree-Fock BCS approach and displayed in Fig. \ref{mean} for SGII, SLy5 and SkM* interactions. The experimental values are also obtained using the three-point mass formula \cite{sat98} and the binding energies given in Refs. \cite{nudat,audi17}. As expected, neutron pairing gap is decreasing, from $^{116}$Sn to $^{130}$Sn, namely from the
mid-shell to the shell closure. With the chosen pairing interaction and strength, our results are compatible with the experimental data.

\begin{table*}
\begin{center}
%\vspace{6mm}
\tabcolsep=0.5em \renewcommand{\arraystretch}{1.0}%
\begin{tabular}
[c]{ccccccccccc}\hline\hline \\ [-1.5ex]
 && \multicolumn{2}{c}{{$^{116}$Sn}}& \multicolumn{2}{c}{{$^{120}$Sn}} & \multicolumn{2}{c}{{$^{124}$Sn}} & \multicolumn{2}{c}{{$^{128}$Sn}}&\\
\hline\\ [-1ex]
 $$& States & $E_{q.p.} $ &$v^{2}$&  $E_{q.p.} $ &$v^{2}$& $E_{q.p.}$ &$v^{2}$&$E_{q.p.}$ &$v^{2}$ \\
\hline\hline\\ [-1.5ex]
&$\nu 1g_{7/2}$ &1.52 & 0.65& 1.77&0.79 & 2.07 &0.89 & 2.38&0.95 \\
&$\nu 3s_{1/2}$&1.22&0.61&1.30 &0.75 &1.41 &0.85& 1.55& 0.93  \\
&$\nu 2d_{3/2}$&1.26&0.32&1.15 &0.51 &1.15 &0.71& 1.23 & 0.88  \\
&$\nu 1h_{11/2}$& 1.65&0.22&1.41 &0.36 &1.24 &0.54& 1.12 &0.75  \\
&$\nu 2f_{7/2}$&6.62&0.01&6.05&0.01 &5.49 &0.01& 4.94 & 0.01  \\
\hline
\end{tabular}
\end{center}
\caption{The quasiparticle energies ($E_{q.p.}$ in MeV) and occupation probabilities ($v^{2}$) of selected neutron states around the Fermi level using Skyrme type SkM* interaction.} \label{table0}
\end{table*}

\begin{figure}[!thb]
  \begin{center}
\includegraphics[width=1\linewidth,clip=true]{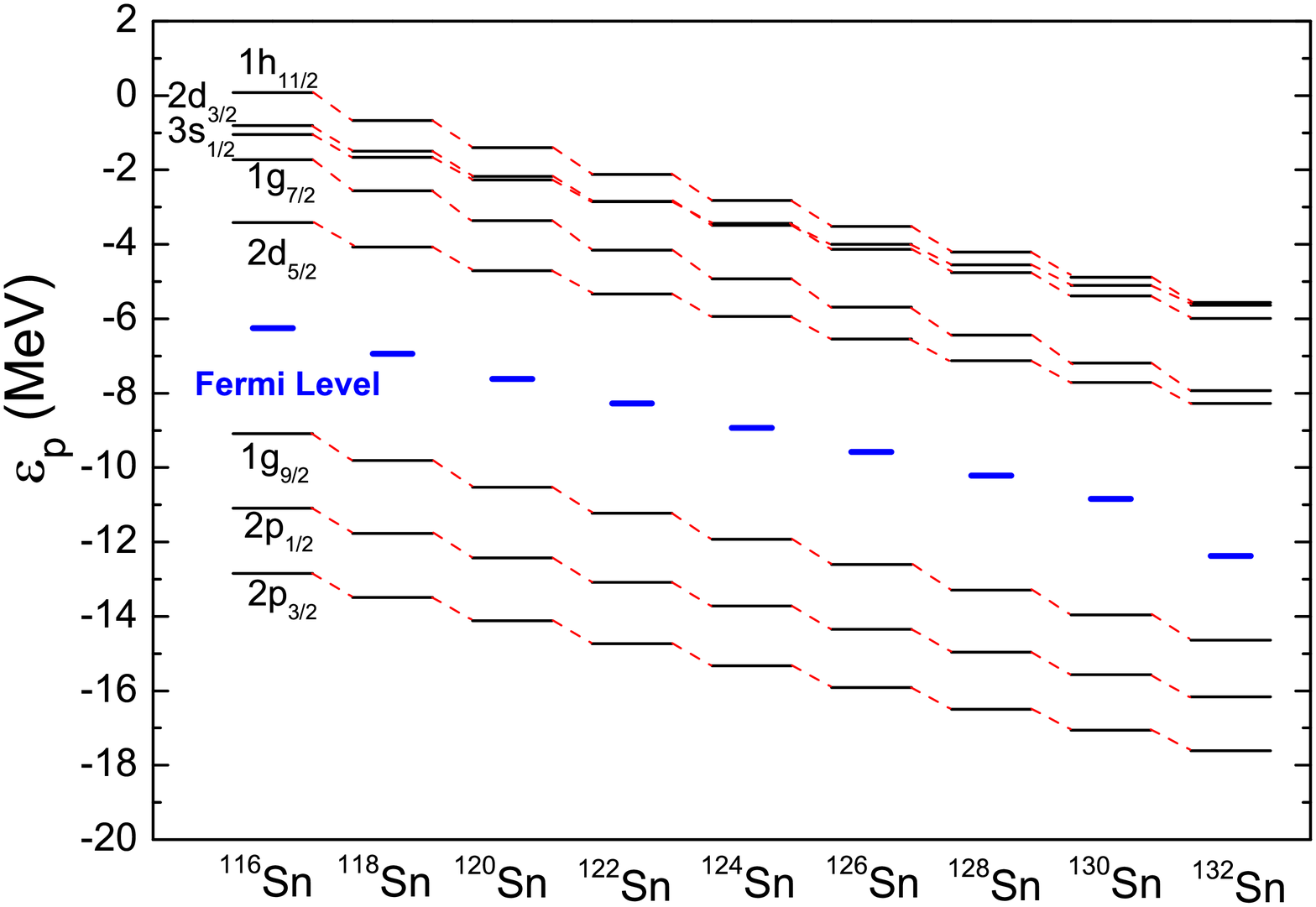}
\includegraphics[width=1\linewidth,clip=true]{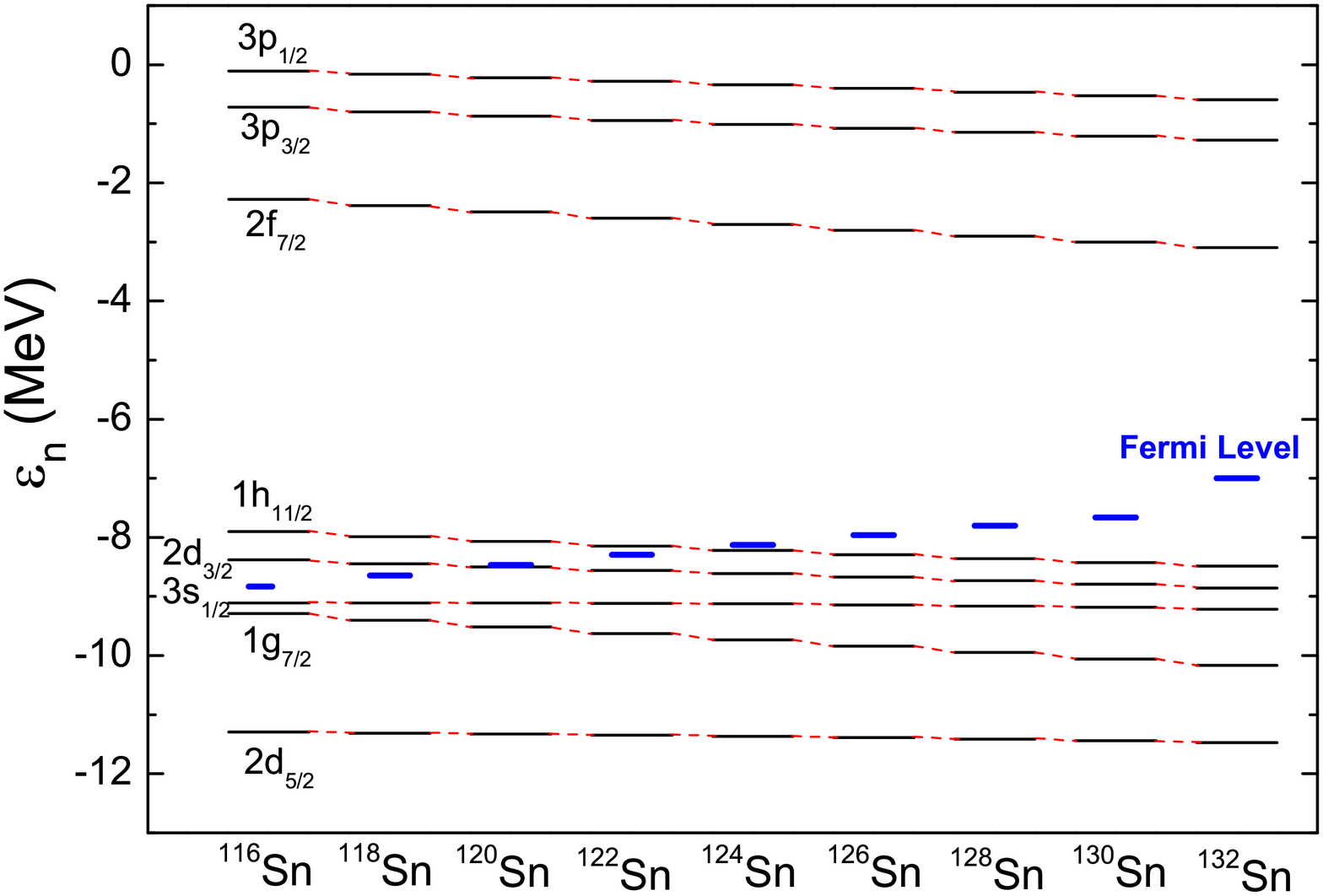}
  \end{center}
 \caption{The proton (upper panel) and neutron (lower panel) single-particle energies ($\varepsilon_{p(n)})$ calculated using SkM* interaction with volume pairing in the Hartree-Fock BCS approach. The Fermi levels are also shown with the thick blue lines (color online).} 
  \label{spe1}
\end{figure}

In  Fig. \ref{spe1}, we also display the proton and neutron single-particle energies for tin nuclei using the Skyrme type SkM* interaction. Both proton and neutron single-particle energies, $\varepsilon_{i}$, 
decrease with increasing neutron number. However, because of the dominance of proton-neutron over neutron-neutron
effective interaction, the effect is stronger on proton energies; in other words,
the proton states are more sensitive to the neutron excess than the neutron 
states, as expected. Therefore, while the proton Fermi energy goes down, 
the neutron Fermi energy (obtained from the Hartree-Fock BCS calculation) moves up with increasing neutron number.

The quasiparticle energy of the neutron states, which is given by 
$E_{i}=\sqrt{(\varepsilon_{i}-\lambda)^{2}+\Delta_{i}^{2}}$, 
is affected by those changes in different ways. In Table \ref{table0}, we 
show the quasiparticle energies and occupation probabilities of selected neutron 
states around the Fermi level for SkM* interaction. 
While the quasiparticle energies of the $1g_{7/2}$ and $3s_{1/2}$  
orbitals increase, they decrease in the case of the $2d_{3/2}$, $1h_{11/2}$ and $2f_{7/2}$ 
orbitals with the increase of the neutron number. For orbitals well above (below) the Fermi energy, this behavior is
expected to be driven by the dominant factor $\vert \varepsilon_i 
- \lambda\vert$ in the quasi-particle energy. For states close to the Fermi energy, the value of
the pairing gap plays a role and the result is more subtle.

As for the $v^2$ factors,
the most important change occurs in the occupation probability of the 
neutron $1h_{11/2}$ state, increasing from 0.22 to 0.75 for 
$^{116}$Sn and $^{128}$Sn, respectively. When it comes to $^{132}$Sn, this state is fully occupied, forming the well known magic number N=82, as expected. 

It should be noted that different Skyrme energy density functionals would predict different single(quasi)-particle energies and occupation probabilities of states. Although we only present here the results using SkM* interaction with the aim of preventing from repetition, the single(quasi)-particle energies and the occupation probabilities exhibit similar behavior with increasing neutron number using SGII and SLy5 interactions.

\subsection{The Isoscalar Quadrupole Response in Superfluid Tin Nuclei} \label{a1}
\subsubsection{Accuracy of the QRPA in the description of the low-energy quadrupole states} \label{a2}
We start our analysis by checking the accuracy of the QRPA in the description of the low-energy quadrupole states in nuclei.

\begin{figure}[!ht]
  \begin{center}
\includegraphics[width=1\linewidth,clip=true]{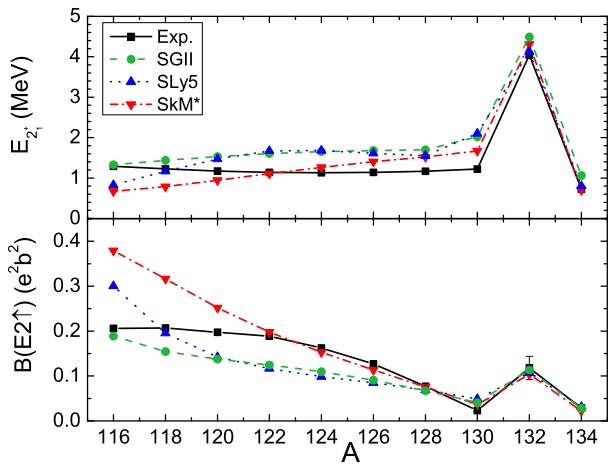}
  \end{center}
 \caption{Upper panel: Experimental and theoretical values of the first $2^+$ state energies for the tin isotopic chain using SGII, SLy5 and SKM* interactions. 
Lower panel: Corresponding reduced electric transition probabilities. The experimental data is taken from Refs. \cite{ra01,prit16}.} 
  \label{spe}
\end{figure}
\begin{figure}[!ht]
  \begin{center}
\includegraphics[width=1\linewidth,clip=true]{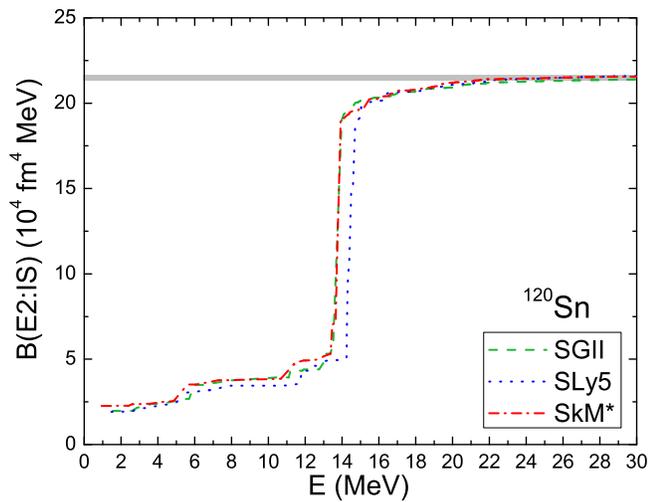}
  \end{center}
 \caption{Running energy-weighted sum for $^{120}$Sn nucleus using SGII, SLy5 and SkM* interactions, respectively. The gray shaded region indicates the range of the double commutator sum rule value for each interaction.}
  \label{run}
\end{figure}
In Figure \ref{spe}, we present the first $2^+$ state energies and the corresponding electric quadrupole transition probabilities together with the experimental data. Both the first $2^+$ state energies and the reduced transition probabilities are well reproduced along the tin isotopic chain within our model calculations, confirming the choice of the pairing interaction.
Although the agreement between our results and experimental data is good, it should be noted that the results also depend on the chosen Skyrme interaction. For instance, SkM* interaction is more successful to reproduce the first $2^+$ state energies and the reduced transition probabilities at the same time for nuclei between $^{120}$Sn and $^{130}$Sn. For lighter nuclei, SGII and SLy5 interactions give better results. This difference comes from the different spectral properties of 
protons and neutrons \cite{fle04}. Pairing predicted by different Skyrme interaction is also known to be quite important for the first $2^+$ state. Since the SkM* interaction is more successful to reproduce the experimental data for heavier tin nuclei, we continue our analysis with SkM* interaction in the rest of this work.

\begin{figure*}[!ht]
 \includegraphics[width=0.47\linewidth,clip=true]{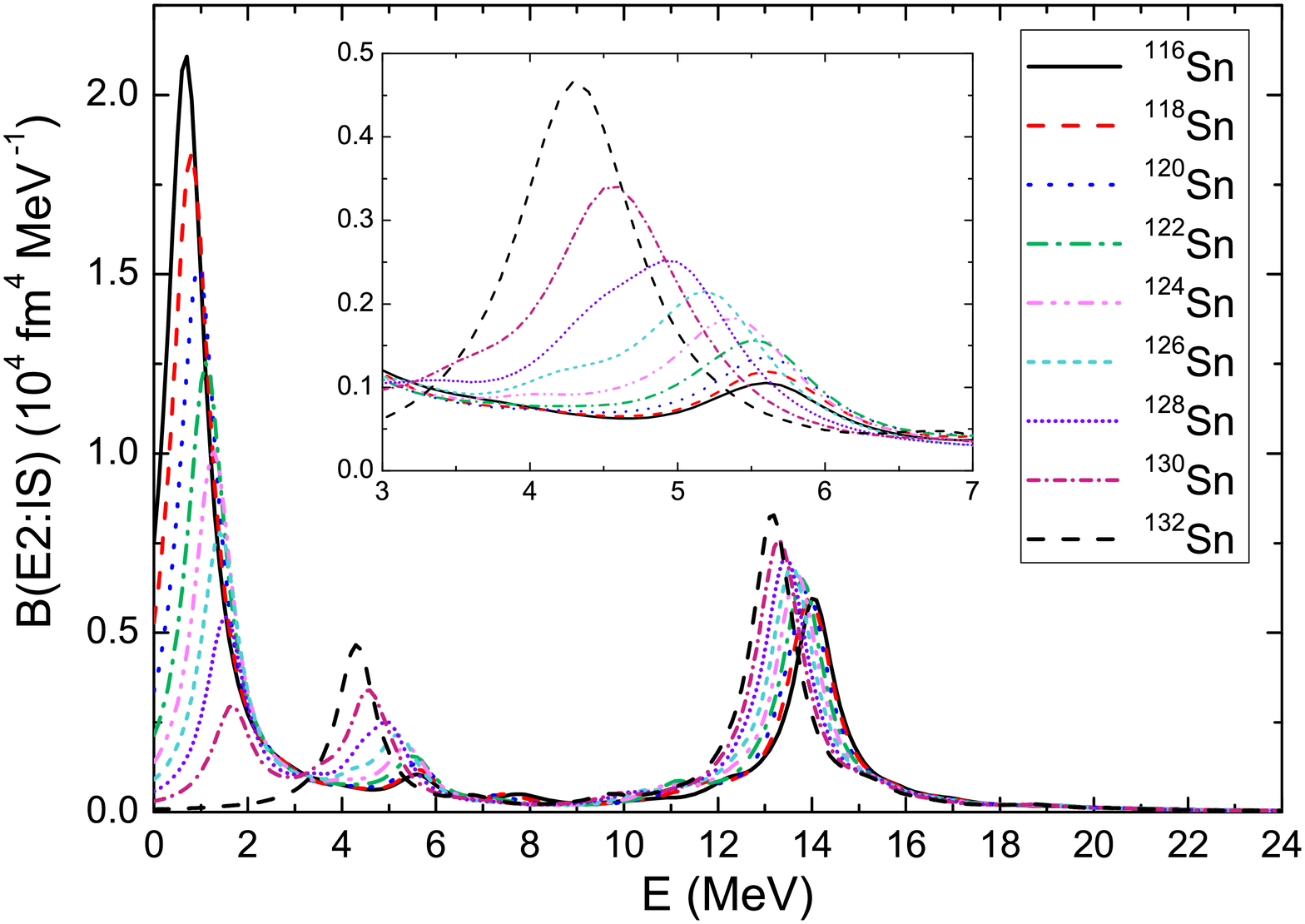}
 \includegraphics[width=0.47\linewidth,clip=true]{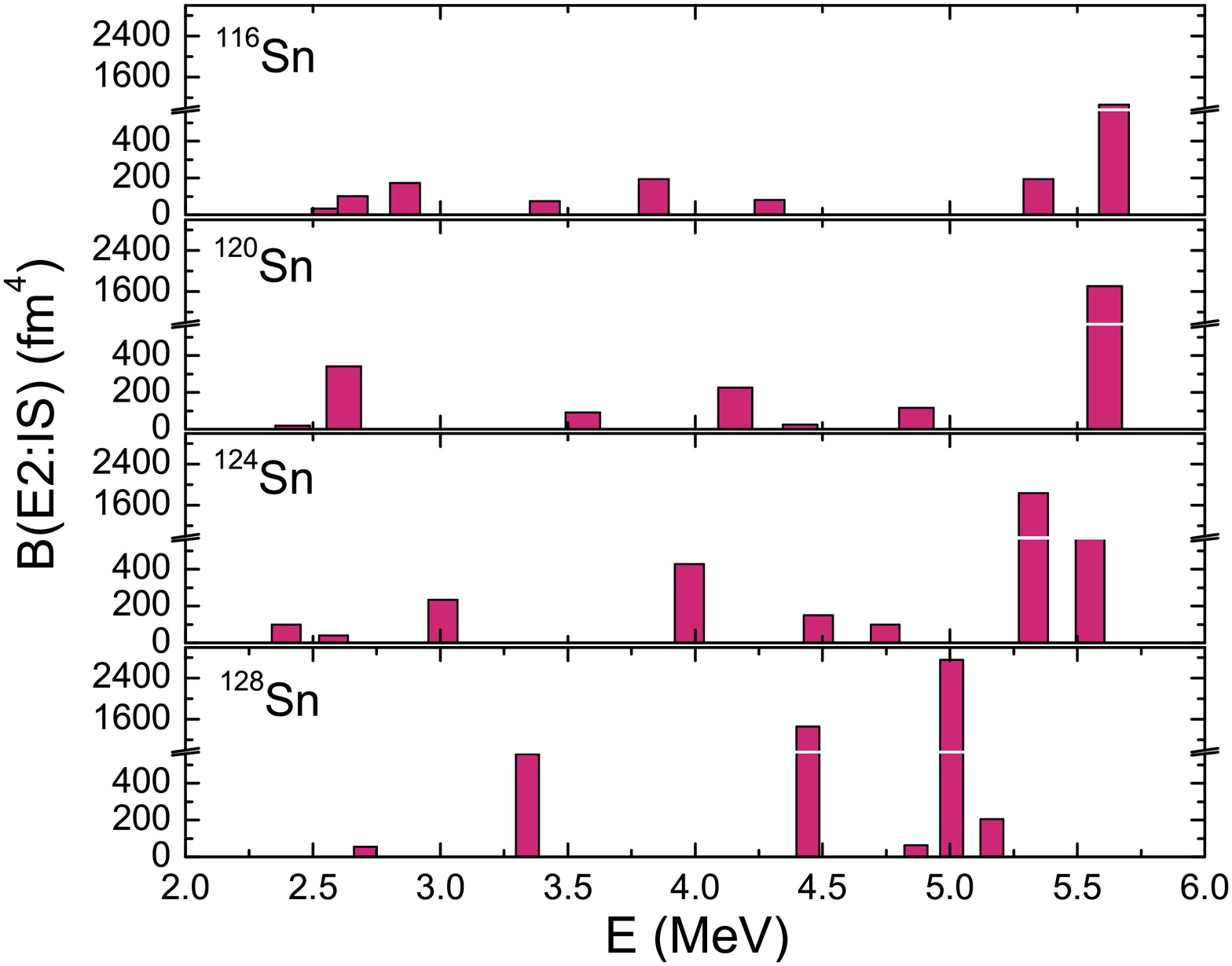}
\caption{Left panel: The isoscalar quadrupole strength in tin nuclei, using SkM* interaction. The low-energy strengths between 3 and 7 MeV are also displayed in the insertion. Right panel: The reduced transition probabilities for the low-energy region of the isoscalar quadrupole response.} 
  \label{9}
\end{figure*}
Let us now study the quasiparticle configurations in the first $2^+$ states of tin nuclei using SkM* interaction.
From $^{116}$Sn to $^{130}$Sn, the contribution of the proton $1g_{9/2}-2d_{5/2}$ transition to the first $2^+$ states gradually decreases from 7.20\% to 2.63\%. Since the reduced electric transition probability depends on the proton excitations, it also gradually decreases (see the lower panel of Fig. \ref{spe}). When it comes to $^{132}$Sn, the contribution of the proton $1g_{9/2}-2d_{5/2}$ transition increases to 24.35\%, leading to an increase in the reduced electric transition probability. The increasing contribution of the proton transition is related to the vanishing of the neutron pairing at N=82 shell closure \cite{sev08}. Finally, it decreases again in $^{134}$Sn, since the contribution of the proton transition is less than 1.0\%. Similar results are also obtained using SGII and SLy5 interactions.

In figure \ref{run}, we also display the running energy-weighted sum for $^{120}$Sn nucleus in order to check for the reliability of the calculations in the isoscalar quadrupole channel. Up to 30 MeV, the isoscalar sum rule saturates and exhausts 99.9\% of the EWSR for each interaction.

\subsubsection{Evolution of the low-energy quadrupole states with increasing neutron number}
In the left panel of Figure \ref{9}, the isoscalar quadrupole strength is displayed 
for $^{116-132}$Sn nuclei using SkM* interaction. In the high-energy 
region above 10 MeV, the isoscalar giant quadrupole resonance states shift slightly 
downward with increasing neutron number. From $^{116}$Sn to $^{132}$Sn, the centroid energies ($E_{c}=m_{1}/m_{0}$) between 10 and 20 MeV decrease smoothly by about 0.7 MeV from 14.07 to 13.36 MeV. Using SkM* interaction, the ISGQR centroid energies roughly follow $68.3A^{-1/3}$ relation, which is consistent with the $64A^{-1/3}$ empirical relation \cite{hara01}.

We also notice an interesting pattern in the low-energy region of the isoscalar quadrupole strength. 
First, the strength of the lowest 2$^+$ state decreases and its energy 
increases from $^{116}$Sn to $^{130}$Sn, as already seen in Fig. \ref{spe}. Second, the low-energy region below the ISGQR
becomes fragmented in heavier neutron-rich nuclei, and some excited states with higher strength 
start to appear between 3 and 6 MeV with the increase of the neutron number.
This pattern can also be seen in the right panel of Fig. \ref{9}, where the reduced isoscalar transition probabilities for the selected tin nuclei are displayed between 2 and 6 MeV. This behavior changes when it comes to the doubly magic $^{132}$Sn nucleus: 
there, the lowest 2$^+$ shows up as a strong peak at 4.3 MeV as can be seen from the left panel of Fig. \ref{9}.

As mentioned above the low-energy quadrupole states between 2 and 5 MeV is named as pygmy quadrupole resonance region in Refs. \cite{pel15,spi16,tso11}. However, a pygmy resonance should correspond to a non-negligible fraction of strength, concentrated in a narrow region, and not due to a trivial threshold effect but to some decoupling of surface neutrons that vibrate independently. 
Along this line, we cannot, in the present work, positively conclude about the existence of a pygmy quadrupole resonance as we will discuss below. Therefore, the states between 2 and 6 MeV are just named as low-energy quadrupole states in the present work.

\begin{table*}
\begin{center}
%\vspace{6mm}
\tabcolsep=0.5em \renewcommand{\arraystretch}{1.0}%
\begin{tabular}
[c]{cccccccccccc}\hline\hline \\ [-1.5ex]
 && \multicolumn{3}{c}{{$^{120}$Sn}} & \multicolumn{3}{c}{{$^{124}$Sn}} & \multicolumn{3}{c}{{$^{128}$Sn}}&\\
\hline\\ [-1.5ex]
 %& &\multicolumn{2}{c}{E=1.18 MeV}&\multicolumn{2}{c}{E=1.18 MeV} &\multicolumn{2}{c}{E=1.18 MeV}& \multicolumn{2}{c}{E=1.18 MeV}\\
%\hline
 $$& Configurations& $E_{conf.}$ &\%& $b_{cd}$& $E_{conf.}$&\%& $b_{cd}$& $E_{conf.}$&\%& $b_{cd}$ \\
 %$$& &MeV & &fm2 &MeV & &fm2 &MeV &  &fm2 \\
\hline
 & $\nu 1h_{11/2}-\nu 1h_{11/2}$&2.82 &26.25 &22.32 &2.48& 44.0 & 25.66 &2.24 & 70.52 & 25.53  \\
& $\nu 1g_{7/2}-\nu 2d_{3/2}$ & 2.93&16.88 & 10.27 &3.22& 9.28&5.58 &3.61&3.47 & 2.09   \\
First $2^+$ & $\nu 3s_{1/2}-\nu 2d_{3/2}$&2.46 &15.92& 8.19 & 2.56 &12.71 &5.48 & 2.77 & 6.0 & 2.41   \\
%\hline
 state & $\nu 2d_{3/2}-\nu 2d_{3/2}$ & 2.31&12.27 & 7.67 &2.29 & 13.88 & 6.48 & 2.45 & 8.33 & 3.36 \\
 & $\nu 1g_{7/2}-\nu 1g_{7/2}$& 3.55 &10.48 &9.05 &4.14& 5.62 & 4.59 &4.76 &2.56 & 1.88   \\
 & $\pi 1g_{9/2}-\pi 2d_{5/2}$& 5.82&6.18 &11.69 &5.98& 4.68 &8.88 &6.16& 2.63&6.08 \\
\hline\\ [-1.5ex]
 %& &\multicolumn{2}{c}{E=1.18 MeV}&\multicolumn{2}{c}{E=1.18 MeV} &\multicolumn{2}{c}{E=1.18 MeV}& \multicolumn{2}{c}{E=1.18 MeV}\\
%\hline
 Low-energy
states at& & &3.55 MeV& & &3.97 MeV& & &4.44 MeV&  \\
\hline
& $\nu 1g_{7/2}-\nu 1g_{7/2}$ &3.55 & 75.90&-15.51 &4.14 &79.10&-12.62  &4.76  &63.86 &-7.81  \\
& $\nu 1g_{7/2}-\nu 2d_{3/2}$ &2.93&4.30   &3.02   &3.21 &4.87  &2.74  &3.61&4.95 &1.93 \\
& $\nu 1h_{11/2}-\nu 1h_{11/2}$ &2.82 &17.66 &11.91& 2.48& 7.55 &7.66  & 2.24&4.74& 4.93 \\
& $\nu 2d_{5/2}-\nu 3s_{1/2}$ &4.40 & 0.90&-1.14& 4.82&2.66  &-1.60  & 5.26&4.35& -1.39 \\
& $\nu 2d_{5/2}-\nu 2d_{3/2}$ &4.25 & && 4.56&2.06  &-0.99  & 4.93&5.03& -1.05 \\
& $\nu 1h_{11/2}-\nu 2f_{7/2}$ &7.86 & && 6.74& 0.70 & -2.12 & 6.06&7.30& -6.82 \\
\hline\\ [-1.5ex]
 %& &\multicolumn{2}{c}{E=1.18 MeV}&\multicolumn{2}{c}{E=1.18 MeV} &\multicolumn{2}{c}{E=1.18 MeV}& \multicolumn{2}{c}{E=1.18 MeV}\\
%\hline
 Low-energy
states at& & &5.60 MeV& & &5.33 MeV& & &5.0 MeV&  \\
\hline
 & $\pi 1g_{9/2}-\pi 2d_{5/2}$& 5.82 &67.20 &-16.57 &5.98& 32.66 & -11.78 &6.16& 24.27& 10.31 \\
 & $\nu1h_{11/2}-\nu 2f_{7/2}$&7.46 &10.54 & -5.60 &6.74& 19.48 & -9.03 & 6.06  & 35.50  & 14.03\\
& $\nu 1g_{7/2}-\nu 1g_{7/2}$& 3.55 &2.74&2.14 &4.14&3.58&2.19 &4.76&20.0 & -4.04  \\
& $\nu 2d_{5/2}-\nu 3s_{1/2}$ &4.40 &2.61 &1.58 & 4.82& 3.45 &1.57 & 5.25&3.06 & 1.21\\
& $\nu 2d_{5/2}-\nu 1g_{7/2}$& 4.87 &3.35 &0.57 &5.48&30.67 &-1.45 & 6.08  &    &  \\
& $\nu 2d_{5/2}-\nu 2d_{3/2}$ &4.25 &0.91 &0.63 & 4.56&  &  & 4.93& 6.90 & -1.15 \\
\hline\hline
\end{tabular}
\end{center}
\caption{The quasiparticle configurations giving the major contribution for the first $2^+$ and the selected low-energy states 
in tin nuclei. For each transition, configuration energies ($E_{conf.}$ in MeV), their contribution to the norm of the state (in percentage) and the corresponding reduced transition amplitudes $b_{ph}$ ($fm^{2}$) (see Eq. \ref{bel}) are given for $^{120}$Sn, $^{124}$Sn and $^{128}$Sn, respectively. Herein, the superscripts $\pi$ and $\nu$ refer to the proton and neutron states, respectively.} \label{table}
\end{table*}
 
\subsubsection{The first $2^+$ state} \label{first}
First, we will explore the underlying mechanism for the decreasing strength of the first 
$2^+$ state. For this purpose, the quasiparticle configurations and their contribution 
to selected excited states are analyzed for the $^{120}$Sn, $^{124}$Sn, and $^{128}$Sn nuclei in Table \ref{table}.
It is known that both the number of the configurations and the coherence among them 
play an important role to determine the strength of an excited state.
The reduced transition amplitudes ($b_{cd}$) for the first $2^+$ state 
show that all configurations provide a contribution having
the same sign, and thus contribute coherently (see Table \ref{table}). 
In the $^{120}$Sn nucleus, several transitions with comparable $b_{cd}$ 
values contribute coherently and the first $2^+$ state exhibit 
large collectivity and strength. 
However, this pattern changes with increasing neutron number. 
The neutron $1h_{11/2}^2$ transition becomes dominant, 
with 70.52\% probability, in $^{128}$Sn (this value reaches 87.90\% in 
$^{130}$Sn). Thus, the collectivity and strength of the first $2^+$ state decreases in heavy tin nuclei.
In the $^{132}$Sn nucleus, the neutron $1h_{11/2}^2$ transition does not contribute to the first $2^+$ state due to the shell closure at N=82. For this nucleus, the neutron $1h_{11/2}-2f_{7/2}$ transition becomes the first possible transition after the shell closure and contributes to the excited state with 64.4\% probability. As mentioned above, the contribution of the proton $1g_{9/2}-2d_{5/2}$ transition to this state also increases to 24.35\%. Therefore, the location of the first $2^+$ state changes and moves to the higher energies.

In order to understand this decrease in the collectivity and strength, let us now 
focus on the two-q.p. configuration energies, and on the occupation probabilities 
of the associated states. The first thing to notice is the change in the 
two-q.p. configuration energies with increasing neutron number.
The configuration energies increase with neutron number, except for the neutron $1h_{11/2}^2$ 
configuration, where it decreases from 2.82 to 2.24 MeV. There is an inverse correlation between the two-q.p. configuration energies and their contribution to the lowest state: if the configuration energies decrease (increase), 
their partial reduced transition amplitudes increase (decrease).
In addition, the occupation probabilities of the states around the Fermi level 
considerably change with neutron number. The matrix elements of the ph and pp interaction do not change much, by themselves, as
a function of the neutron number.
Nonetheless, for the ph interaction, the $u$ and $v$ factors [cf. Eqs. (\ref{eq:a}) and (\ref{eq:b})] 
are responsible for keeping the ph matrix elements non-negligible for the neutron $1h_{11/2}^2$ configuration, and
decreasing the ph matrix elements between that configuration and the others. This is due to the change on the occupation probabilities of the neutron states with increasing neutron number, as explained above in Section \ref{1}. As a consequence, the neutron $1h_{11/2}^2$ transition decouples from other transitions, and gives the most important contribution to the first $2^+$ excited state in heavier and open-shell tin isotopes. Due to the decreasing strength of the attractive residual interaction, the first $2^+$ excited state energies also increase with increasing neutron number.

\subsubsection{The low-energy states}

\begin{table*}
\begin{center}
%\vspace{6mm}
\tabcolsep=0.5em \renewcommand{\arraystretch}{1.0}%
\begin{tabular}
[c]{ccccccccccc}\hline\hline \\ [-1.5ex]
  \multicolumn{3}{c}{{$^{120}$Sn}} & \multicolumn{3}{c}{{$^{124}$Sn}} & \multicolumn{3}{c}{{$^{128}$Sn}}&\\
\hline\\ [-1.5ex]
 %& &\multicolumn{2}{c}{E=1.18 MeV}&\multicolumn{2}{c}{E=1.18 MeV} &\multicolumn{2}{c}{E=1.18 MeV}& \multicolumn{2}{c}{E=1.18 MeV}\\
%\hline
  E (MeV) &B(E2:IS) &\% EWSR&   E (MeV) &B(E2:IS) &\% EWSR&   E (MeV) &B(E2:IS)&\% EWSR \\
 %$$& & &0.95 MeV & & &1.26 MeV & & &1.52 MeV&  \\
\hline
 2.62 &342.0&0.41 &3.01&243.4& 0.31 &3.34 &557.2 &0.78 \\
 3.55  & 91.0&0.15& 3.97 &426.7& 0.74 &4.44  &1459.3&2.71    \\
 4.15 &225.5&0.43& 5.33  & 1835.7 &4.28 &5.0  &2761.8& 5.78   \\
%\hline
 5.60 &1712.0&4.43& 5.55 &704.8 & 1.72 & 5.16 & 203.8& 0.44  \\
\hline\\ [-1.5ex]
\end{tabular}
\end{center}
\caption{The selected excited states in the low-energy region of the isoscalar quadrupole strength, corresponding reduced transition probabilities (in units of $fm^{4}$), and their contribution to the EWSR is presented in percentage for $^{120}$Sn, $^{124}$Sn and $^{128}$Sn nuclei, respectively.} \label{table2}
\end{table*}
 
Pygmy dipole or quadrupole states have not been unambiguously characterized in the previous literature. If, in the low-energy sector of the spectrum, some concentrated and non-negligible 
fraction of strength is found we could talk about a pygmy state, provided it is clearly decoupled from other states. In this case, it is also implicitly assumed that it is mainly a neutron state and likely to be concentrated on the nuclear surface.

The recent theoretical results obtained with QPM on top of the HFB-QRPA indicate the presence of some low-energy excited states in neutron-rich tin nuclei, that have been named as pygmy quadrupole resonance states \cite{tso11,pel15,spi16}. Using the QPM, it was found that the low-energy states between 2 and 5 MeV exhaust 4.1\% of the isoscalar EWSR for $^{124}$Sn nucleus \cite{spi16}. Experimental data are also available for $^{124}$Sn nucleus and the excitation energies below 5 MeV \cite{pel15,spi16}. Between 2 and 5 MeV, the low-energy states exhaust 3.8(5)\% and 5.5(6)\% of the isoscalar EWSR for the ($\alpha,\alpha'$) and nuclear resonance fluorescence (NRF) \cite{spi16, en15} data, respectively. Using QRPA, the quadrupole states exhaust 0.62\% of the EWSR in $^{124}$Sn nucleus, which is quite low compared to the experimental data and the QPM calculations. Nonetheless, we find excited states just above 5 MeV, exhausting a considerable amount of the EWSR and that may correspond to the missing EWSR in the low-energy region of $^{124}$Sn nucleus. Including the excited states at 5.33 and 5.55 MeV, quadrupole states exhaust 4.26\% of the EWSR, which is comparable with the QPM and the experimental data.
In addition, we find that the transition probabilities of the isoscalar quadrupole states increase with increasing neutron number, which is consistent with the results given in Ref. \cite{tso11}. It should be mentioned that the QRPA usually predicts a strength distribution which is reasonable as a whole, while the coupling with more complicated configurations changes the precise values of the energies and the details of the strength fragmentation. Therefore, one may need to vary the excitation energy interval in the low-energy region in order to make comparison with the QPM results and experimental data. However, we believe that the qualitative nature of our conclusions is robust in this manner.
Although one-by-one comparison with the experimental data is not possible in the low-energy region within the QRPA framework, the underlying reasons for the increase of the isoscalar quadrupole strength with neutron number can be clarified. 
Therefore, in the present work, we analyze the structure and evolution of these states with increasing neutron number.
Apart from the most visible peaks above 5 MeV on Fig. \ref{9}, we also consider the excited states below 5 MeV with small transition probabilities. 

In Table \ref{table2}, we present the results for the selected low-energy states with corresponding reduced transition probabilities using SkM* interaction. Their contribution to the EWSR is also given in percentage. By increasing neutron number, several excited states start to appear in the low-energy region with considerable transition probabilities.
While the excited states start to accumulate between 3 and 6 MeV, their strength and contribution to the EWSR increases. Considering their behavior, a further investigation of these low-energy quadrupole states is relevant to check for their nature.

To start with, we find that the second and third $2^+$ states in neutron-rich tin nuclei lie below 3 MeV, except for doubly magic $^{132}$Sn nucleus. In $^{120}$Sn, $^{124}$Sn and $^{128}$Sn nuclei, second and third $2^+$ states have quite low strength and are found mainly formed with the neutron $2d_{3/2}-2d_{3/2}$ and $3s_{1/2}-2d_{3/2}$ transitions, respectively. Similar results were also obtained in Refs. \cite{tso11,spi16}.

In order to clarify the underlying mechanism of the increase of the isoscalar strength between 3 and 6 MeV, we analyze in the case of the prominent states with considerable transition probabilities, their major components and the associated contribution to the transition probabilities in Table \ref{table}. As we mentioned above, apart from the states with the highest transition probabilities at 5.60, 5.33 and 5.0 MeV in $^{120}$Sn, $^{124}$Sn, and $^{128}$Sn, respectively, we also analyze some excited states below 5 MeV with comparable strengths. 
In Table \ref{table} we also present these low-energy states at 3.55, 3.97 and 4.44 MeV for $^{120}$Sn, $^{124}$Sn, and $^{128}$Sn, respectively. By analyzing these states, we obtain cancellations between the $b_{cd}$ transition amplitudes, at variance with the first $2^+$ states. These excited states are mainly formed with one or two two-q.p. excitations and do not show a clear collective behavior. In addition, these states are neutron dominated, and mainly based on transitions within the same sub-shell. With increasing neutron number, the excited states start to take contribution from various neutron transition channels that appear in $^{128}$Sn nucleus. Nonetheless, the collectivity of these states do not change much with neutron number. 

On the contrary, the states above 5 MeV are mainly made from several excitation channels and have higher transition probabilities, compared to the states below 5 MeV. By increasing neutron number, the neutron transitions 
also start to play a dominant role in this region: the strength increases and the energy shifts down gradually. Similar to the states below 5 MeV, we also find cancellations in the reduced transition amplitudes. 
In addition, we obtain an interesting behavior in this region. First, the lowest proton $1g_{9/2}-2d_{5/2}$ and neutron $1h_{11/2}-2f_{7/2}$ transitions between the shell gaps play a major role for these states.
Second, these states change their behavior, switching their structure from proton dominated to neutron dominated excited states with increasing neutron number. 
For instance, in the $^{116}$Sn nucleus, we obtain an excited state at 5.64 MeV with 1065.0 $fm^{4}$, exhausting 2.90\% of the isoscalar EWSR (see fig. \ref{9}). This state is almost formed with one two-q.p. configuration: proton $1g_{9/2}-2d_{5/2}$ (76.05\%). In the proton channel, the $1g_{9/2}-2d_{5/2}$ transition corresponds to the first possible transition after Z=50 gap.
By increasing the neutron number, the contribution of the proton $1g_{9/2}-2d_{5/2}$ transition gradually decreases (see Table \ref{table}).
Meanwhile, the contribution of neutron transitions start to increase, leading to an increase in the isoscalar strength. 

\begin{figure*}[!th]
  \begin{center}
\includegraphics[width=1.\linewidth,clip=true]{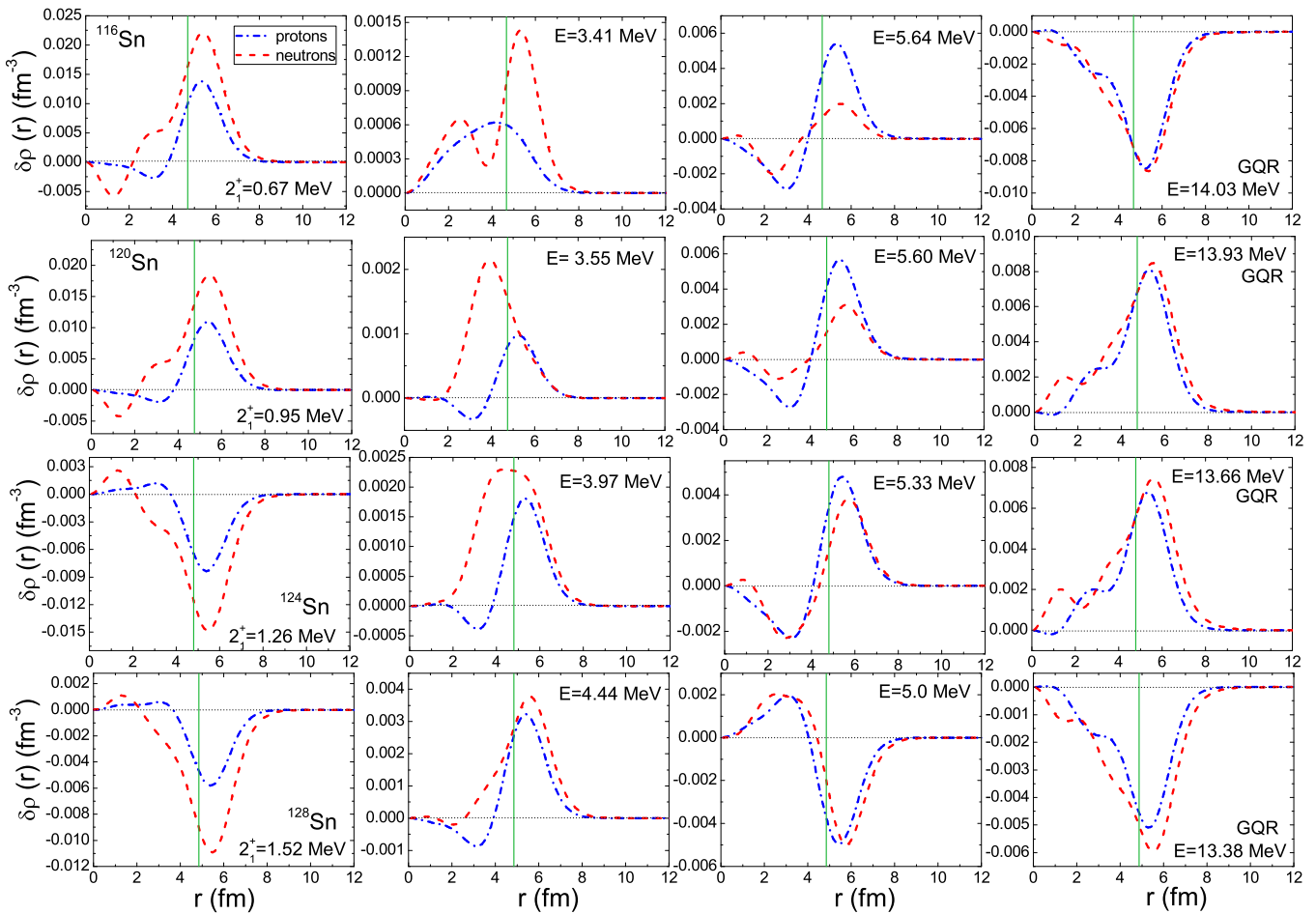}
  \end{center}
  \caption{Proton and neutron transition densities for selected states in $^{116}$Sn, $^{120}$Sn, $^{124}$Sn, and $^{128}$Sn nuclei. The leftmost and the rightmost columns represent the transition densities for the first $2^+$ states and GQR regions, respectively. In the middle, the proton and neutron transition densities are shown for the most prominent peaks in the low-energy region. The vertical green lines (color online) represent the calculated neutron radii for each nuclei.} 
  \label{td}
\end{figure*}

The second important contribution to the excited states above 5 MeV comes from neutron $1h_{11/2}-2f_{7/2}$ transition. This transition also corresponds to the N=82 gap for tin chain. In contrast to the proton $1g_{9/2}-2d_{5/2}$ transition, 
the contribution of the neutron $1h_{11/2}-2f_{7/2}$ transition increases with increasing neutron number. 
For instance, from $^{120}$Sn to $^{128}$Sn, the contribution of this transition increases from 10.54\% to 35.50\%, and reaches 38.0\% 
in $^{130}$Sn. In order to explain the increasing contribution of this transition, we analyze the unperturbed two-q.p. energies and the residual interaction with increasing neutron number. The configuration energies are increasing with neutron number, except the neutron $1h_{11/2}-2f_{7/2}$ transition. 
From $^{120}$Sn to $^{128}$Sn, the two-q.p. energy of this transition decreases from 7.46 to 6.06 MeV.  
Moreover, as mentioned above in Sec.\ref{first}, the change in the occupation probabilities of the states with increasing neutron number affects the magnitude of the attractive residual interaction: it increases for the neutron $1h_{11/2}-2f_{7/2}$ transition.
Since the occupation probability of the neutron $2f_{7/2}$ state barely changes with increasing neutron number, the filling of the $1h_{11/2}$ state has an important impact on the neutron $1h_{11/2}-2f_{7/2}$ transition, leading to an increase of the residual interaction, the reduced transition amplitude and the isoscalar strength, respectively. 
In addition, we find that the attractive isoscalar residual interaction becomes weaker for other two-q.p. configurations with increasing neutron number. 
Under the influence of a weak residual interaction, their location do not change much. Therefore, only two-q.p. configuration transitions from $2d_{5/2}-1g_{7/2}$ (for $^{124}$Sn) and $1g_{7/2}-1g_{7/2}$ (for $^{128}$Sn), which lie close to these excitation energies around 5 MeV, can mix with the proton $1g_{9/2}-2d_{5/2}$ and neutron $1h_{11/2}-2f_{7/2}$ transitions in heavy tin nuclei.

Considering the structure of the low-energy excited states, we conclude that the main reason of the increase in the low-energy strength is the increasing contribution of neutron transitions, due to the filling of the high-$j$ neutron $1h_{11/2}$ and $1g_{7/2}$ states with the increase of the neutron number. While the excited states below 5 MeV are sensitive to the excitations within the same sub-shell, the excited states above 5 MeV could also provide information about the shell evolution of neutron-rich nuclei, since the most important transitions in the low-energy region correspond to the proton Z=50 and neutron Z=82 shell gaps. Therefore, fragmentation of the isoscalar quadrupole strength in the low-energy region can be related with the changing shell structure of nuclei with increasing neutron number. 

The present calculations can be also performed using different Skyrme energy density functional parameters 
in order to check the interaction dependence of the results. Although the SkM* interaction is used in the calculations due to its success in the description of the experimental data for the first $2^+$ state energies and the corresponding electric 
quadrupole transition probabilities of heavier tin nuclei, we obtain similar results using Skyrme type SLy5 and SGII interactions. We also find that in the Pb isotopic chain, from $^{200}$Pb to $^{208}$Pb, 
a similar trend emerges: while the isoscalar quadrupole strength decreases for the first $2^+$ state, some excited states start to accumulate between 3 and 6 MeV, whose strength increases with increasing neutron number.

\subsubsection{Transition Densities}
The transition densities are also useful in order to clarify the nature and behavior of these low-energy states. 
An excited state can be purely isovector (IV)
if the transition densities of protons and neutrons are out of phase. On the contrary, 
an excited state can be defined as purely isoscalar (IS) if
the neutron and proton transition densities display in phase motion.

In figure \ref{td}, proton and neutron transition densities are displayed for the selected low-energy quadrupole states in $^{116}$Sn, $^{120}$Sn, $^{124}$Sn, and $^{128}$Sn nuclei. In the first $2^+$ state, transition densities display an isoscalar behavior, with dominant neutron contribution in the surface region. Similar to the first $2^+$ state, protons and neutrons also oscillate in phase in the GQR region, as expected.
In the middle two columns, we display the transition densities for the most important peaks above and below 5 MeV. As stated above, these states are sensitive to the neutron excess, increasing their strength with increasing neutron number. In contrast to the first $2^+$ state and the isoscalar giant quadrupole state transition densities, we find that the states below 5 MeV with lower transition probabilities display a different behavior in $^{120}$Sn, $^{124}$Sn, and $^{128}$Sn nuclei. Although the protons and neutrons mainly oscillate in phase, we also obtain a mixture of isoscalar and isovector motion of protons and neutrons inside the nuclei. In addition, the neutron transition densities start to shift through the surface region with increasing neutron number, while the proton transition densities are not affected. This behavior is in agreement with the features of the transition densities of those states between 2 and 4 MeV, which are referred as the pygmy quadrupole type oscillation of nuclei in Refs. \cite{tso11,spi16}. In $^{116}$Sn, the transition densities display a different behavior compared to the heavier tin nuclei. We obtain an isoscalar motion of protons and neutrons with dominant neutron contribution through the surface region. This difference in the transition densities is related with the difference in the two-q.p. configurations of the corresponding excited states. While neutron $1g_{7/2}-1g_{7/2}$ is the main two-q.p. configuration for the excited states in heavier tin nuclei, both neutron $1g_{7/2}-1g_{7/2}$ (44.50\%) and $1h_{11/2}-1h_{11/2}$ (52.77\%) two-q.p configurations are obtained in $^{116}$Sn nucleus.

In addition, the states above 5 MeV with higher strength display an isoscalar behavior.
The proton transition densities are dominant up to $^{120}$Sn. By increasing neutron number and after $^{124}$Sn nucleus, the neutron transition densities start to contribute more through the surface region. This is also an indication for the change from proton dominated excitation to the neutron one.

The pygmy dipole resonance, which is rooted in the neutron excess of nuclei, may be described as an oscillation of the neutron skin against the isospin saturated proton-neutron core \cite{paar09}. It is also known that the properties of the pygmy dipole resonance region are correlated with the neutron skin thickness of nuclei \cite{pie11,bar13}. The low-energy quadrupole states also display some similarities as compared to the pygmy dipole resonance properties, namely, their strength increase and excitation energies decrease with increasing neutron number between 3 and 6 MeV. However, transition densities do not display strong neutron dominance in the surface region compared to the pygmy dipole resonance \cite{roc12,yuk12}. The neutron transition densities start to be apparent in the surface region only after the $^{124}$Sn nucleus. Therefore, we believe that the fragmentation pattern in the low-energy region occurs due to the changes in the shell structure of nuclei with increasing neutron number, and the low-energy states do not display a quadrupole-type skin core oscillation of nuclei.

\section{Summary and Conclusions} \label{finito}
In the present work, we performed QRPA calculations for the isoscalar quadrupole response of tin nuclei. The Skyrme type SGII, SLy5 and SkM* interactions are used in the calculations with volume type pairing. The first $2^+$ states and the corresponding electric reduced transition probabilities of the states are well reproduced in our calculations.

We investigate the changes in the low-energy region of the isoscalar quadrupole strength with increasing neutron number. The low-energy region of the isoscalar quadrupole strength is fragmented with increasing neutron number. While the first $2^+$  state strength decreases, some low-energy states start to accumulate between 3 and 6 MeV, increasing their strength with the neutron number. 

Below 5 MeV, the excited states are neutron dominated and mainly display a non-collective character. Transitions between the same sub-shells play an important role in this region, and the summed transition strength between 2 and 5 MeV increases with neutron number. In addition, the transition densities of these states display a specific pattern. In neutron-rich tin nuclei, the proton and neutron transition densities 
exhibit a mixture of isoscalar and isovector character inside the nuclei, and an isoscalar motion of protons and neutron is obtained through the surface, which is different from the first $2^+$ state and GQR behavior. In addition, neutron transition densities start to dominate through the surface of nuclei with increasing neutron number.
Our results are also consistent with the findings in Ref. \cite{tso11,spi16}. 
Nonetheless, we cannot positively conclude about the existence of a pygmy quadrupole resonance, and 
we deem that more experimental and theoretical works are needed in order to clarify the nature of these states in the low-energy region.

The low-energy states above 5 MeV are also affected by the increase of the neutron number. While the collectivity and strength of these states increase, their excitation energies shift downwards with increasing neutron number.
We also find that these states change their structure from proton dominated to neutron dominated excited states with the increase of the neutron number. The configurations of these states show that they are sensitive to the proton and neutron shell gaps, and could also provide information about the shell evolution in open-shell nuclei. The transition densities also show that the protons and neutrons display an isoscalar character. 

From our analysis, we conclude that the fragmentation of the low-energy strength is related with the changes of shell structure of nuclei with increasing neutron number. Especially, the filling of the high-$j$ neutron $1h_{11/2}$ and $1g_{7/2}$ levels has an important impact for the increase of the low-energy strength between 3 and 6 MeV. 
Similar calculations are also performed with different Skyrme energy density functionals and nuclei, and we find that the fragmentation of the isoscalar quadrupole strength in the low-energy region is independent from the chosen interaction and isotopic chain.

\section{Acknowledgments} 
This work is supported by the Croatian Science Foundation under the project Structure and Dynamics
of Exotic Femtosystems (IP-2014-09-9159). E. Y. acknowledges financial support from the Scientific
and Technological Research Council of Turkey (T\"{U}B\.{I}TAK) BIDEB-2219 Postdoctoral Research program. Funding from the European Union's Horizon 2020 research and innovation programme under grant agreement No 654002 is also acknowledged.


\begin{thebibliography}{99}

\bibitem{paar00} N. Paar, D. Vretenar, E. Khan, and G. Col\`{o}, Rep. Prog. Phys. \textbf{70}, 691 (2007).

\bibitem{hara01} M.N. Harakeh and A. van der Woude, \textit{Giant Resonances} (Oxford University Press, Oxford, 2001).

\bibitem{paar09} N. Paar, Y.F. Niu, D. Vretenar, and J. Meng, Phys. Rev. Lett. \textbf{103}, 032502 (2009).

\bibitem{vre12} D. Vretenar, Y.F. Niu, N. Paar, and J. Meng, Phys. Rev. C \textbf{85}, 044317 (2012).

\bibitem{roc12} X. Roca-Maza, G. Pozzi, M. Brenna, K. Mizuyama, and G. Col\`{o}, Phys. Rev. C \textbf{85}, 024601 (2012).

\bibitem{lan09} E.G. Lanza, F. Catara, D. Gambacurta, M.V. Andr\`{e}s, and Ph. Chomaz, Phys. Rev. C \textbf{79}, 054615 (2009).

\bibitem{yuk12} E. Y\"{u}ksel, E. Khan, and K. Bozkurt, Nucl. Phys. A \textbf{877}, 35 (2012).

\bibitem{sav13} D. Savran, T. Aumann, and A. Zilges, Prog. Part. Nucl. Phys. \textbf{70}, 210 (2013).

\bibitem{bra16} A. Bracco, F.C.L. Crespi, and E.G. Lanza, Eur. Phys. J. A \textbf{51}, 99 (2015).

\bibitem{wie09} O. Wieland, A. Bracco, F. Camera, G. Benzoni, N. Blasi, S.
Brambilla, F. C. L. Crespi, S. Leoni, B. Million, R. Nicolini, A.
Maj, P. Bednarczyk, J. Grebosz, M. Kmiecik, W. Meczynski, J.
Styczen, T. Aumann, A. Banu, T. Beck, F. Becker, L. Caceres,
P. Doornenbal, H. Emling, J. Gerl, H. Geissel, M. Gorska, O.
Kavatsyuk, M. Kavatsyuk, I. Kojouharov, N. Kurz, R. Lozeva,
N. Saito, T. Saito, H. Schaffner, H. J. Wollersheim, J. Jolie, P.
Reiter, N. Warr, G. deAngelis, A. Gadea, D. Napoli, S. Lenzi, S.
Lunardi, D. Balabanski, G. LoBianco, C. Petrache, A. Saltarelli,
M. Castoldi, A. Zucchiatti, J. Walker, and A. B\"{u}rger, Phys. Rev. Lett. \textbf{102}, 092502 (2009).

\bibitem{krum15} A.M. Krumbholz, P. von Neumann-Cosel, T. Hashimoto, A. Tamii, T. Adachi, C.A. Bertulani, H. Fujita, Y. Fujita, E. Ganioglu, K. Hatanaka, C. Iwamoto, T. Kawabata, N.T. Khai, A. Krugmann, D. Martin, H. Matsubara, R. Neveling, H. Okamura, H.J. Ong, I. Poltoratska,  V.Yu. Ponomarev, A. Richter, H. Sakaguchi, Y. Shimbara, Y. Shimizu, J. Simonis, F.D. Smit, G. Susoy, J.H. Thies, T. Suzuki, M. Yosoi, and J. Zenihiro, Phys. Lett. B \textbf{744}, 7 (2015).

\bibitem{gori02} S. Goriely and E. Khan, Nucl. Phys. A \textbf{706}, 217 (2002).

\bibitem{lit09} E. Litvinova, H.P. Loens, K. Langanke, G. Mart\`{\i}nez-Pinedo, T. Rauscher, P. Ring, F.-K. Thielemann, and V. Tselyaev, Nucl. Phys. A \textbf{823}, 26 (2009).

\bibitem{pie06} J. Piekarewicz, Phys. Rev. C \textbf{73}, 044325 (2006).

\bibitem{klim07} A. Klimkiewicz, N. Paar, P. Adrich, M. Fallot, K. Boretzky, T. Aumann, D. Cortina-Gil, U. Datta Pramanik, Th. W. Elze, H. Emling, H. Geissel, M. Hellstr\"{o}m, K.L. Jones, J.V. Kratz, R. Kulessa, C. Nociforo, R. Palit, H. Simon, G. Sur\'{o}wka, K. S\"{u}mmerer, D. Vretenar, and W. Walu\'{s}, Phys. Rev. C \textbf{76}, 051603(R) (2007).

\bibitem{tar97} O. Tarasov, R. Allatt, J.C. Ang\`{e}lique, R. Anne, C. Borce, Z. Dlouhy, C. Donzau, S. Gr\`{e}vy, D. Guillemaud-Mueller, M. Lewitowicz, S. Lukyanov, A.C. Mueller, F. Nowacki, Yu. Oganessian, N.A. Orr, A.N. Ostrowski, R.D. Page, Yu. Penionzhkevich, F. Pougheon, A. Reed, M.G. Saint-Laurent, W. Schwab, E. Sokol, O. Sorlin, W. Trinder and J.S. Winfield,  Phys. Lett. B \textbf{409} 64 (1997).

\bibitem{yok95} M. Yokoyama, T. Otsuka, N. Fukunishi, Phys. Rev. C \textbf{52} 1122 (1995).

\bibitem{ham97} I. Hamamoto, H. Sagawa, and X.Z. Zhang, Phys. Rev. C \textbf{55}, 2361 (1997).

\bibitem{lan03}  K. Langanke, J. Terasaki, F. Nowacki, D.J. Dean, and W. Nazarewicz, Phys. Rev. C \textbf{67}, 044314 (2003).

\bibitem {tso11} N. Tsoneva and H. Lenske, Phys. Lett. B \textbf{695}, 174 (2011).

\bibitem {pel15} L. Pellegri, A. Bracco, N. Tsoneva, R. Avigo, G. Benzoni,
N. Blasi, S. Bottoni, F. Camera, S. Ceruti, F.C.L.
Crespi, A. Giaz, S. Leoni, H. Lenske, B. Million, A.I.
Morales, R. Nicolini, O. Wieland, D. Bazzacco, P. Bednarczyk,
B. Birkenbach, M. Ciema la, G. de Angelis,
E. Farnea, A. Gadea, A. G\"{o}rgen, A. Gottardo, J. Grebosz,
R. Isocrate, M. Kmiecik, M. Krzysiek, S. Lunardi,
A. Maj, K. Mazurek, D. Mengoni, C. Michelagnoli, D.R.
Napoli, F. Recchia, B. Siebeck, S. Siem, C. Ur, and J.J.
Valiente-Dob\`{o}n, Phys. Rev. C, \textbf{92}, 014330 (2015).

\bibitem {spi16} M. Spieker, N. Tsoneva, V. Derya, J. Endres, D. Savran, M.N. Harakeh, S. Harissopulos, R.-D. Herzberg, A. Lagoyannis, H. Lenske, N. Pietralla, L. Popescu, M. Scheck, F. Schlüter, K. Sonnabend, V.I. Stoica, H.J. W\"{o}rtche, and A. Zilges, Phys. Lett. B \textbf{752}, 102 (2016).

\bibitem{ter02} J. Terasaki, J. Engel, W. Nazarewicz, M. Stoitsov, Phys. Rev. C \textbf{66}, 054313 (2002).

\bibitem{fle04} P. Fleischer, P. Kl\"{u}pfel, P.-G. Reinhard, and J A. Maruhn, Phys. Rev. C \textbf{70}, 054321 (2004).

\bibitem{ter06} J. Terasaki and J. Engel, Phys. Rev. C \textbf{74}, 044301 (2006).

\bibitem{an06} A. Ansari and P. Ring, Phys. Rev. C \textbf{74}, 054313 (2006).

\bibitem{sev08} A.P. Severyukhin, V.V. Voronov, and N.V. Giai, Phys. Rev. C \textbf{77}, 024322 (2008).

\bibitem{scam13} G. Scamps and D. Lacroix, Phys. Rev. C \textbf{88}, 044310 (2013).

\bibitem{niu16} Y.F. Niu, G. Col\`{o}, E. Vigezzi, C.L. Bai, and H. Sagawa, Phys. Rev. C \textbf{94}, 064328 (2016).

\bibitem{bar82} J. Bartel, P. Quentin, M. Brack, C. Guet, and H.-B. H\aa kansson, Nucl. Phys. A \textbf{386}, 79 (1982).

\bibitem{SGII} N.V. Giai and H. Sagawa, Phys. Lett. B \textbf{106}, 379 (1981).

\bibitem{SLY5} E. Chabanat, P. Bonche, P. Haensel, J. Meyer, and R. Schaeffer, Nucl. Phys. A \textbf{635}, 231 (1998).

\bibitem{ber91} G.F. Bertsch and H. Esbensen, Ann. Phys. \textbf{209}, 327 (1991).

\bibitem{sat98} W. Satu\l{}a, J. Dobaczewski, and W. Nazarewicz, Phys. Rev. Lett. \textbf{81}, 3599 (1998).

\bibitem{ring80} P. Ring, P. Schuck, \textit{The Nuclear Many-Body Problem} (Springer, Berlin, 1980).

\bibitem{suho07} J. Suhonen, \textit{From Nucleons to Nucleus: Concepts of Microscopic Nuclear Theory} (Springer, Berlin, 2007).

\bibitem{li08} J. Li, G. Col\`{o}, and J. Meng, Phys. Rev. C \textbf{78}, 064304 (2008).

\bibitem{bai14} C.L. Bai, H. Sagawa, G. Col\`{o}, Y. Fujita, H.Q. Zhang, X.Z. Zhang, and F.R. Xu, Phys. Rev. C \textbf{90}, 054335 (2014).

\bibitem{paar03} N. Paar, P. Ring, T. Nik\v{s}i\`{c}, and D. Vretenar, Phys. Rev. C \textbf{67}, 034312 (2003).

\bibitem{niu17}  Z.M. Niu, Y.F. Niu, H.Z. Liang, W.H. Long, and J. Meng, Phys. Rev. C \textbf{95}, 044301 (2017). 

\bibitem{nudat} National Nuclear Data Center, information extracted from the NuDat 2 database, http://www.nndc.bnl.gov/nudat2/

\bibitem{audi17} G. Audi, F.G. Kondev, M. Wang, W.J. Huang, and S. Naimi, Chin. Phys. C \textbf{41}, 030001 (2017).

\bibitem{ra01} S. Raman, C.W. Nestor, P. Tikkanen, At. Data Nucl. Data Tables, \textbf{78}, 1 (2001).

\bibitem{prit16} B. Pritychenko, M. Birch, B. Singh, M. Horo, At. Data Nucl. Data Tables, \textbf{107}, 1 (2016).

\bibitem{en15} ENSDF, NNDC online data service, ENSDF database, http://www.nndc.bnl.gov/ensdf/, (2015).

\bibitem{pie11} J. Piekarewicz, Phys. Rev. C \textbf{83}, 034319 (2011).

\bibitem{bar13} V. Baran, M. Colonna, M. Di Toro, A. Croitoru, and D. Dumitru, Phys. Rev. C \textbf{88}, 044610 (2013).

\end{thebibliography}
\end{document}